\documentclass[10pt]{IEEEtran}

\IEEEoverridecommandlockouts                              % This command is only needed if 
\usepackage{svg}
\usepackage[english]{babel}
\usepackage[utf8]{inputenc}
\usepackage{lipsum}
\usepackage[cmex10]{amsmath}
\usepackage{amssymb}
\usepackage{xcolor}
\usepackage{color}
\usepackage{float}
\usepackage{tikz}
\usepackage{multirow}
\usepackage{makecell}
\usepackage{graphicx}
\usepackage[linesnumbered,ruled,boxed]{algorithm2e}
\renewcommand{\KwData}{\textbf{Input:}}  % Use Input in the format of Algorithm  
\renewcommand{\KwResult}{\textbf{Output:}} % Use Output in the format of Algorithm
\usepackage{algcompatible}
\usepackage{cite}
\usepackage{url}
\usepackage{array}

{}
\newtheorem{corollary}{Corollary}{}
{}
{}
\newtheorem{conjecture}[corollary]{Conjecture}
{}
{}

\newcommand{\m}{\boldsymbol}
\DeclareMathOperator{\Tr}{trace}
\DeclareMathOperator{\diag}{diag}
\renewcommand{\baselinestretch}{1}

\title{\LARGE \bf Enhancing Vehicle Platooning Safety via Control Node Placement and Sizing under State and Input Bounds}
\author{Yifei She$^{1}$, Shen Wang$^{*1}$, Ahmad Taha$^{2}$, and Xiaofeng Tao$^{1}$ \thanks{$^{1}$National Engineering Research Center of Mobile Network Technologies, Beijing University of Posts and Telecommunications, Beijing 100876, China \{bupt3.1415926, shen.wang, taoxf\}@bupt.edu.cn.}% 
	\thanks{$^{2}$Department of Civil and Environmental Engineering, Vanderbilt University, Nashville, TN, USA. {ahmad.taha@vanderbilt.edu}.} 
	\thanks{*Shen Wang is the corresponding author.}
	\thanks{This work is supported by the NSFC under grants No. 62203062 and No. 61932005, the Fundamental Research Funds for the Central Universities under Grant 2024RC08, and NSF under grant No. 2152450.}}

\begin{document}
	\maketitle
	\begin{abstract}

		Vehicle platooning with Cooperative Adaptive Cruise Control improves traffic efficiency, reduces energy consumption, and enhances safety but remains vulnerable to cyber-attacks that disrupt communication and cause unsafe actions. To address these risks, this paper investigates control node placement and input bound optimization to balance safety and defense efficiency under various conditions. We propose a two-stage actuator placement and actuator saturation approach, which focuses on identifying key actuators that maximize the system's controllability while operating under state and input constraints. By strategically placing and limiting the input bounds of critical actuators, we ensure that vehicles maintain safe distances even under attack. Simulation results show that our method effectively mitigates the impact of attacks while preserving defense efficiency, offering a robust solution to vehicle platooning safety challenges. 
		%\begin{IEEEkeywords}
		%Water distribution networks, state estimation, geometric programming, linearization. 
		%\end{IEEEkeywords}
		
	\end{abstract}
	\section{Introduction and Paper Contributions}

\begin{figure} 
		\centering
		\includegraphics[width=0.75\linewidth]{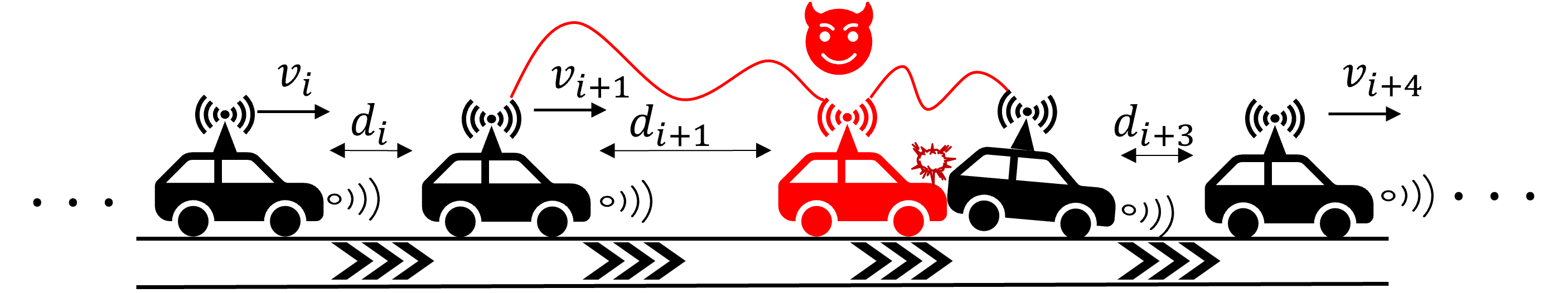}
		%\caption{An illustrative vehicles platooning scenario with $n$ vehicles. Each vehicle is equipped with a CACC strategy and detects adjacent vehicles using radar. The $i$-th vehicle's speed is $v_i$, and the distance from the car in front is denoted as $d_{i}$. An attack can hijack vehicle communication to manipulate the input $u_i$ of the secondary control strategy.}
		\caption{A vehicle platooning scenario with $n$ vehicles, each using CACC and radar to detect adjacent vehicles. The $i$-th vehicle's speed is $v_i$, and the distance to the preceding vehicle is $d_{i}$. An attack can hijack communication to manipulate the input $u_i$ of the secondary control.}
		\label{fig:VP}
	\end{figure}

	In modern transportation systems, vehicle platooning has emerged as a promising technology to improve traffic efficiency, reduce energy consumption, and ensure road safety~\cite{goodplatoon}. Utilizing advanced control systems such as Cooperative Adaptive Cruise Control (CACC)~\cite{CACCreview}, vehicles can maintain safe distances and speeds; see Fig.~\ref{fig:VP} for a typical platooning system with $n$ vehicles. However, disruptions from attackers can lead to severe accidents, causing significant damage to life and property. As a result, vehicle platooning is a safety-critical issue that requires continuous effort to maintain its reliability and safety~\cite{defendplatoon1}.

	A key challenge in vehicle platooning is ensuring that system states impacted by throttle and brake actuators remain within safe limits under different conditions. These actuators control the engine or motor to regulate speed, enabling the vehicle to maintain consistent acceleration or deceleration. However, attackers can exploit vulnerabilities in the CACC system. For example, a man-in-the-middle attack~\cite{maninthemiddle} or a jamming attack~\cite{jammingattack} could intercept and manipulate vehicle communication and alter critical data, leading to improper actions such as sudden braking or acceleration. This can disrupt the platooning system's reachable set and push it into unsafe states; see Fig.~\ref{fig:VP} for illustrations. This issue is studied through control theory as a reachability problem, assessing whether the system's reachable set intersects predefined danger zones~\cite{Murguia2020,ActuatorSaturation}.

	%Although MITM attacks may severely impact a platoon, defenders can still deploy defensive strategies to counter such attacks. For example, the authors of \cite{ActuatorSaturation,DARIA} proposed redesigning the physical bounds of all actuators to protect the platooning system. With proper design, the system can remain safe. It is worth noting that while imposing input bounds on all actuators limits the attacker's capabilities, it also reduces the flexibility of the platooning system, meaning that certain states can never be reached, further weakening the system's resilience.

	While attacks can severely impact a platoon, defense strategies can mitigate these risks. For instance, the authors in~\cite{ActuatorSaturation, DARIA} propose redesigning actuator bounds to limit attackers' capacity through actuator saturation, as physical constraints prevent actuators from injecting excessive energy into the system. Although this method enhances safety, it reduces the system's defense efficiency by restricting all actuators, limiting its ability to maintain safety with minimal resource use.
	
	Instead of restricting all actuators' bounds, focusing on selecting a subset of critical actuators with the most significant impact on the system's controllability as defensive positions is more effective. Limiting only these actuators' bounds allows the system's safety and defense efficiency to be maintained and balanced, even in the worst-case scenario where all actuators are compromised. If only some actuators are attacked, this defense strategy offers even greater efficiency, as only the key actuators need to be managed. Therefore, the primary focus of this paper is determining the optimal number, placement, and input bounds for these critical actuators to ensure vehicle platooning safety.
	
	%In fact, we have found that it is not necessary to impose restrictions on all actuators. Instead, it is sufficient to select a few actuators that have the most substantial impact on the controllability of the platooning system and limit their input bounds. Under this defensive strategy, the system's safety and resilience can still be guaranteed even if an MITM attack compromises all actuators. Suppose only some of the actuators are attacked. In that case, this strategy ensures safety and resilience and provides additional flexibility, allowing the reachable set of the platooning system to be more extensive. Therefore, this paper's primary focus is determining the optimal number of actuators and their corresponding placement and redesigning their input bounds to ensure the platooning system's safety.
	
	\subsection{Related Works and Research Problem}
	
	Various studies have focused on actuator saturation in the control theory field to prevent attacks or disturbances that could endanger systems. Studies~\cite{ActuatorSaturation} and~\cite{DARIA} design new physical limits on actuators to deter attackers from arbitrarily manipulating the system, and the authors in~\cite{ConstrainingAttackers,butchkocps2021} go a step further and design limits on actuators such that the reachable set avoids dangerous states and includes desired operation states.  
	The authors in~\cite{Escudero2021} study the possibility of constraining the controller output (i.e., the input to the actuators) utilizing a dynamic filter designed to prevent the reachability of dangerous plant states.  
	Recently, they have considered stealthy actuator and sensor attacks and proposed a set-theoretic method to synthesize safety-preserving filters for closed-loop systems~\cite{Escudero2023}. 
	
	While control theory offers insights into actuator saturation, its application in vehicle platooning still needs to be explored. In real-world contexts, actuators must balance both safety and efficiency. Current research emphasizes the role of actuator saturation in critical scenarios such as emergency braking, where it ensures safe vehicle spacing, and in variable time headway strategies, which improve stability and reduce congestion~\cite{transportation1, transportation2}. These findings underscore the importance of actuator saturation for the robustness and safety of vehicle platooning systems.
	
	% Control theory defines actuator saturation, but its practical applications in transportation systems, particularly vehicle platooning, need further exploration. Translating these theoretical insights into real-world transportation contexts is challenging, where actuators must balance safety and overall efficiency. The current research on actuator saturation in vehicle platooning systems highlights its crucial role under various challenging conditions. In emergency braking scenarios, actuator saturation is essential for ensuring safe spacing between vehicles~\cite{transportation1}. Moreover, under the variable time headway strategy, actuator saturation can help maintain the stability and responsiveness of the platoon, effectively reducing traffic congestion and improving road utilization efficiency ~\cite{transportation2}. Together, these advancements demonstrate the importance of actuator saturation for the robustness and safety of modern vehicle platooning systems.
	
	However, all aforementioned studies~\cite{DARIA, ActuatorSaturation, ConstrainingAttackers, butchkocps2021, Escudero2021, Escudero2023,transportation1,transportation2} only consider actuator saturation and do not take into account how to choose which actuators to control when the number of controllable actuators is limited to secure critical engineering safety since \textit{(i)} the actuator placement problem itself is NP-hard~\cite{ MinimalActuatorPlacement, taha_time-varying_2019} and \textit{(ii)} the coupling of actuator selections and limits/bounds makes securing critical engineering safety a complex and computationally expensive task, and decoupling the impacts of actuator selections and limits on reachability set is challenging. 
	
	\subsection{Paper Contributions and Organization}
	
	Hence, this work's main contributions are constraining the attacker's capabilities and securing vehicle platooning safety through co-optimizing actuator saturation and placement. The key contributions of this paper are as follows:  
	
	\begin{itemize} 
		
		\item  This paper presents a novel defense strategy integrating actuator placement and the corresponding bounds to enhance the safety and defense efficiency of vehicle platooning. Moreover, we transform the system dynamics by decoupling the actuator location and the corresponding bound for independent problem-solving. 
		
		\item  A two-stage algorithm is proposed to constrain the attacker's capabilities. Specifically, the actuator placement stage selects the most controllable locations with a limited number of actuators, while the actuator saturation stage redesigns the corresponding bounds to avoid dangerous states. 
		
		\item  The effectiveness of the proposed method is verified via numerical simulations, demonstrating the improved safety and defense efficiency of the optimized vehicle platooning system. 
	\end{itemize}
	
	The rest of this manuscript is organized as follows. The problem statement formulation is given in Section~\ref{sec:PSF}. Section~\ref{sec:ASAP} proposes actuator placement
	and saturation algorithm aiming to enhance the safety of the overall system. Section~\ref{sec:casestudy} illustrates the main results via numerical simulations. Section~\ref{sec:conclusion} concludes.
	
	\textbf{Notation:}
	$\mathbb{N}$ and $\mathbb{R}$ denote the sets of natural and real numbers. For a set $\mathcal{V}$, its cardinality is $|\mathcal{V}|$. We use $\mathbb{R}^{n}$ and $\mathbb{R}^{m\times n}$ for an $n$-dimensional column vector and an $m \times n$ matrix, respectively. For $\m x \in \mathbb{R}^{n}$, $\m x_i$ is its $i$-th element, and $\m x^{\top}$ is its transpose. In a matrix ${\m A} \in \mathbb{R}^{n \times m}$, $\m A_{ij}$ refers to the element at row $i$, column $j$. Notation ${\m A} \succ {0}$ (${\m A} \succeq {0}$) indicates a positive (semi-) definite matrix. The identity and zero matrices are denoted as $\m I$ and $\m 0$, with dimensions inferred from context.
	%The notations $\mathbb{N}$ and $\mathbb{R}$ denote the set of natural and real numbers, and given a set $\mathcal{V}$, we denote as $|\mathcal{V}|$ its cardinality. We denote $\mathbb{R}^{n},\mathbb{R}^{m\times n}$ as a column vector with $n$ elements and an $m$-by-$n$ matrix in $\mathbb{R}$. For any vector $\m x \in \mathbb{R}^{n}$,  $\m x_i$ is its  $i$-th element, and  $\m x^{\top}$ is its  transpose. For a matrix ${\m A}_{n \times m}$ with $n$ rows and $m$ columns, $\m A_{ij}$ is its element located at the $i$-th row and $j$-th column, and ${\m A} \succ {0}$ (${\m A}\succeq {0}$) means $\m A$ is positive (semi-) definite. We denote the identity matrix and zero matrix as $\m I$ and $\m 0$, whose dimensions are inferred from the context.
	
	\section{Problem Statement and Formulation} \label{sec:PSF}
	\subsection{System and Attack Models Using CACC Strategy}
	We consider a classic vehicles platooning system with $n$ vehicles equipped with actuators enabling control of the acceleration/deceleration of each vehicle, following a leader vehicle closely; see Fig.~\ref{fig:VP} for details. We have the following dynamics using the CACC strategy~\cite{ConstrainingAttackers}
	\begin{equation}
		\begin{aligned}
			%d_{1}(k+1) & =d_{1}(k)+\Delta t[v_{2}(k)-v_{1}(k)] \\
			%& \vdots \\
			d_{i-1}(k+1)  & =d_{i-1}(k)+\Delta t [v_{i}(k)-v_{i-1}(k) ] \\
			%v_{1}(k+1) & =v_{1}(k)+\beta_1 v_{1}(k)+\Delta t \widetilde{u}_{1}(k) \\
			%& \vdots \\
			v_{i}(k+1) & =v_{i}(k)+\beta_i v_{i}(k)+\Delta t \widetilde{u}_{i}(k),
		\end{aligned}
	\end{equation}
	where $d_{i}$ is the distance between the $i$-th and $(i+1)$-th vehicles; $v_i(k)$ is the $i$-th vehicle's speed; $\Delta t$ stands for the sampling period; $\beta_i<0$ characterizes velocity loss attributed to friction for each vehicle $i$, and $\widetilde{u}_i(k)$ symbolizes the control input affecting acceleration/deceleration. To ensure safety, adjacent vehicles must not collide and should strive to maintain the desired distance $d^*_i, i = 1,\ldots,n-1$ and velocity $v^*_i, i = 1,\ldots,n$. 
	
	We consider a CACC strategy combining an adaptive cruise control and a secondary control given by $\widetilde{u}_i(k) = \hat{u}_i(k) + {u}_i(k),$
	where $\hat{u}_i(k)$ is a forward-and-reverse-looking proportional-derivative control~\cite{VehicularPlatooning} to maintain a desired distance by accelerating or decelerating the vehicle, and
	\begin{align*}
		\hat{u}_i(k) =&-k_i^p[d_{i-1}(k)-d_{i-1}^*]  + k_i^p[d_i(k)-d_i^*] \\
		&+ k_i^d[v_{i-1}(k)-v_i(k)] 
		+ k_i^d[v_i(k)-v_{i+1}(k)],
	\end{align*}
	where $k_i^p$ and $k_i^d$ are proportional gain and derivative gain. 
	
	Moreover, ${u}_i(k)$ is the secondary control strategy that relies on a communication network susceptible to jamming attack~\cite{jammingattack} or other types of attacks to transmit system state information and control instructions. %An attacker can manipulate the control input ${u}_i(k)$ by hijacking the network to apply jamming attack~\cite{jammingattack}, disruption attack~\cite{VPDattack} or other types of attacks.
	These attacks can lead to unsafe scenarios, such as reducing the distance between adjacent vehicles below the safety threshold or causing a crash. Because of the physical limits in the acceleration/deceleration of each vehicle, we assume that the secondary control action is bounded by $\underline{u}_i \leq u_{i}(k) \leq \bar{u}_i$\cite{inputsaturationinplatoon}. By changing the upper bound $\bar{u}_i$ and lower bound $\underline{u}_i$ of $u_i(k)$ to constrain the acceleration of the vehicle, it is an essential method for us to resist attacks.
	
	\subsection{System Model Representation and Dangerous Set}
	We introduce the differences between actual and desired distances and velocities as new variables, that are $\Delta d_i = d_i - d_i^*$ and $\Delta v_i = v_i - v_i^*$, and let the state variable  $\m x(k) = [\Delta d_1(k), \cdots, \Delta d_{n-1}(k), \Delta v_1(k), \cdots, \Delta v_{n}(k)]^{\top}$, and the secondary control input variable  $\m u(k) = [u_1(k), \cdots, u_{m}(k)]^{\top}$. Thus, we have a discrete-time Linear Time-Invariant (LTI) system featuring individually bounded control inputs, described by 
	\begin{equation} \label{eq:LTI}
		\m x(k+1) = {\m A} \m x(k) + \m B \m u(k),
	\end{equation}
	where $k \in \mathbb{N}$; ${\m x}(k) \in \mathbb{R}^{2n-1}$ denotes the state; ${\m A} \in \mathbb{R}^{(2n-1) \times (2n-1)}$ is the state matrix; ${\m B} \in \mathbb{R}^{(2n-1) \times m}$ represents the $m$ inputs, and note that the none-zeros in ${\m B}$ have two meanings: \textit{(i)} the location of actuators and \textit{(ii)} the amplification factor of the system input $\m u$; $\m u(k) \in \mathbb{R}^m$ stands for symmetrically bounded control input such that
	\begin{equation} \label{eq:actuatorbounds}
		-\sqrt{\gamma_i} = \underline{u}_i \leq	  u_i(k) \leq \bar{u}_i = \sqrt{\gamma_i},\quad i=1,\dots,m,
	\end{equation} 
	where $\gamma_i>0$ is a constant determining the bound for each control input, and $u_i(k)>0$ ($u_i(k)<0$) indicates the vehicle acceleration (deceleration). %Here, $\m x_i$ denotes the $i^\text{th}$ element of $\m{x}$.
	The actuator bounds are collected in the matrix $\m \Gamma$, and $\m \Gamma:= \diag(\frac{1}{\gamma_1},\ldots,\frac{1}{\gamma_m}).$
	
	In summary, the reachable set $\mathcal{R}$ of LTI system $\eqref{eq:LTI}$ from the initial state $\m x(0)$ with actuator locations and corresponding bounds are
	\begin{equation}
		{\mathcal{R}}:=\left\{\m x(k) \in \mathbb{R}^n \left\lvert\, \begin{array}{l}
			\hspace{-0.6em} \m x(k+1) = {\m A} \m x(k) + \m B \m u(k) \\
			{ \m x(0)=\m 0, u_i^2(k) \leq {\gamma}_i}
		\end{array}\right.\right\}.
	\end{equation}
	
	%In this paper, we note that the size of the reachable set indicates system flexibility. Flexibility in this vehicle platooning system refers to the system's ability to maintain a wide range of possible states (i.e., a sizeable reachable set) while ensuring safety.
	
	We define the dangerous set $\mathcal{D}$ as
	\begin{equation} \label{eq:danger}
		\mathcal{D} := \{\m x\in \mathbb{R}^n\ |\  \cup_{i=1}^{\kappa}\m c_i^\top \m x \geq \m b_i \},
	\end{equation}
	where each $\m c_i^\top \m x \geq {\m b}_i$ represents a half-space. Any vector in $\mathcal{D}$ signifies the failure to maintain safe distances and speeds, necessitating the redesign of actuator bounds to prevent such a situation.
	
	The actuator locations and bounds are pivotal factors affecting the system's reachable set $\mathcal{R}$, ultimately determining the system's ability to avoid entering a dangerous state $\mathcal{D}$. We can select a proper number of actuators and redesign the corresponding bounds to ensure $ \mathcal{R} \cap \mathcal{D} = \emptyset$. Through the optimization of the actuator locations and input bounds, it becomes possible to enhance the safety of the vehicle platooning system while selecting minimum numbers of actuators and maintaining defense efficiency.

	\subsection{System Model Transformation}\label{sec:decoupling}
	From system model~\eqref{eq:LTI}, we can see that both actuator locations and bounds significantly impact the final reachable set. That is, both input matrix $\m B$ and control input $\m u$ in \eqref{eq:LTI} induce reachable set $\mathcal{R}$, but in a coupled way, i.e., $\m B \m u$. 
	
	We assume the sets $\mathcal{V}$ and $\mathcal{S}$ contain all possible locations and selected actuator locations. It is clear that $|\mathcal{S}| \leq |\mathcal{V}|= m$. We decouple their impact on system states by assuming the input matrix is only composed of zeros or ones, i.e., only encodes the actuator locations (denoted as $\m B_{\mathcal{S}}$), and thus the actuator bounds are taken care of by control input ${\m u}_\mathcal{S}$. That is, $\m B_{\mathcal{S}} \in \mathbb{R}^{(2n-1) \times m}$ is a matrix such that if the actuator location is selected in $\mathcal{S}$, the corresponding entry of $\m B$ is one; otherwise it is zero. Similarly, each entry in ${\m u}_\mathcal{S}$ equals the corresponding entry in ${\m u}$ multiplied by a coefficient from the original $\m B$. That is, $\m B \m u  = (\m B_{\mathcal{S}}\m E) \m u=\m B_{\mathcal{S}} (\m E \m u) =\m B_{\mathcal{S}} {\m u}_\mathcal{S}$, where $\m E \in \mathbb{R}^{m\times m}$ stores the coefficients. In the end, $\m B \m u$ turns into $\m B_{\mathcal{S}} {\m u}_\mathcal{S}$, and the final system model of the vehicle platooning system is
	\begin{equation} \label{eq:LTIS}
		\m x(k+1) = {\m A} \m x(k) + \m B_{\mathcal{S}}{\m u}_\mathcal{S}(k),
	\end{equation}
	where $\m B_{\mathcal{S}}$ only encodes actuator locations and ${\m u}_\mathcal{S}$ only encodes actuator bounds as claimed. If all actuator locations are selected (i,e., $\mathcal{S} = \mathcal{V}$), we have
	%Considering a $n$-vehicle scenario in Fig.~\ref{fig:VP} and using the CACC strategy, we give the system and input matrices in~\eqref{eq:LTIS} when all locations are placed with actuators (i,e., $\mathcal{S} = \mathcal{V}$). Therefore,
	\begin{equation}\label{equ:systemmatrix}
		\small \begin{matrix}
			\m A = \begin{bmatrix}
				\m I_{n-1} & \m A^{(12)}_{(n-1) \times n}\\
				\m A^{(21)}_{n \times (n-1)} &   \m A^{(22)}_{n \times n}
			\end{bmatrix},&
			\m B _{\mathcal{S}}  = \begin{bmatrix}
				\m 0_{(n-1) \times n}\\
				\m I_{n}
			\end{bmatrix}
		\end{matrix},
	\end{equation}
	where the submatrices in $\m A$ are 
	\begin{align*}
		\m A^{(12)} & = \scriptsize{\begin{bmatrix}
				-1 & 1 & \ldots & 0 \\ 
				\vdots & \ddots & \ddots & 0 \\ 
				0 & \ldots & -1 & 1 
			\end{bmatrix} \hspace{-2pt} \Delta t,} {\normalsize \m A^{(21)}}   = \scriptsize \begin{bmatrix}
			k_{1}^p & \ldots & 0 \\ 
			-k_{2}^p & k_{2}^p & 0 \\ 
			\vdots & \ddots & \ddots \\ 
			0 & \ldots & -k_{n}^p 
		\end{bmatrix}, \\ 
		\m A^{(22)} & = \scriptsize \begin{bmatrix}
			(1+\beta_1)-k_{1}^d & k_{1}^d & 0 & 0 \\ 
			k_{2}^d & (1+\beta_2)-2k_{2}^d & k_{2}^d & 0 \\ 
			\vdots & \ddots & \ddots & \vdots \\ 
			0 & \ldots & k_{n}^d & (1+\beta_n)-k_{n}^d 
		\end{bmatrix}.
	\end{align*}

	Note that the reachable set $\mathcal{R}$ of LTI system $\eqref{eq:LTI}$ equals the reachable set $\mathcal{R}_\mathcal{S}$ of LTI system $\eqref{eq:LTI}$, that is, $\mathcal{R} = \mathcal{R}_\mathcal{S}$ since $\m B \m u  =\m B_{\mathcal{S}} {\m u}_\mathcal{S}$.
	
	\subsection{Actuator Saturation and Actuator Placement Problem}
	The Actuator Saturation and Actuator Placement (\text{\textbf{ASAP}}) defense strategy aims to enhance the safety and defense efficiency of vehicle platooning by \textit{(i)} selecting a small fraction of the actuators maximizing the controllability and \textit{(ii)} redesigning their bounds. Mathematically, the defense strategy can be formulated as an optimization problem
	\begin{subequations} \label{eq:ASAP-new}
		\begin{align}
			\text{\textbf{ASAP:}} \; \;  \;\underset{\mathcal{S},\hat{\m \Gamma}}{\text{min}}  & \quad   g(\mathcal{S},\hat{\m \Gamma})  \\
			\text{s. t.}  &\quad \mathcal{S} \subseteq \mathcal{V}, |\mathcal{S}| \leq m,\\
			& \quad  \hat{\m \Gamma} \geq \m \Gamma,  \label{equ:newboundasap} \\
			& \quad  \mathcal{R}_\mathcal{S} \cap \mathcal{D} = \emptyset \label{equ:safety}
		\end{align} 
	\end{subequations}
	where the constant $m$ is the number of actuator locations (i.e., system inputs) to defense in a vehicle platooning system set up by system operators due to defense cost, the newly designed actuator bounds $\hat{\m \Gamma}$ should be smaller than the original $\m \Gamma$ (i.e., constraint \eqref{equ:newboundasap} holds) to ensure the safety (i.e., constraint \eqref{equ:safety}). The objective function $g(\mathcal{S},\hat{\m \Gamma})$ is a function of actuator set $\mathcal{S}$ and the corresponding bounds $\hat{\m \Gamma}$ to be redesigned, and we choose it as the maximization of controllability of the LTI system representation of~\eqref{eq:LTIS}.
	
	This problem is a set-function optimization problem that involves both discrete and continuous variables. The selection of the actuator set $\mathcal{S}$ is a discrete variable, representing a binary decision for each potential actuator location, i.e., whether an actuator is selected to defend against attack. Meanwhile, the actuator bounds $\hat{\m \Gamma}$ are continuous variables, representing the continuous redesign of the bounds for the chosen actuators. Combining these discrete and continuous variables creates a vast search space due to the exponential growth with respect to $\mathcal{S}$. This combination significantly increases the computational complexity, as the solver must explore both binary set selections and continuous permutations.
	
	It is challenging to solve the \text{\textbf{ASAP}} problem directly because of the undefined mathematical representation of reachable set $\mathcal{R}_\mathcal{S}$. To directly overcome the difficulty of describing a reachable set $\mathcal{R}_\mathcal{S}$, we can choose an approximate method to formulate a reachable set, greatly reducing the computation cost and obtaining a feasible solution.
	
	\section{Enhance Safety of Vehicle Platooning by a two-stage ASAP algorithm} \label{sec:ASAP}
	In this section, we propose a two-stage algorithm to solve the \text{\textbf{ASAP}} problem~\eqref{eq:ASAP-new}. Initially, we involve the optimal actuator locations in determining the approximate shape of the reachable set for the defense strategy. Then, by solving an optimization problem, we determine the input bounds and scale the sizes of the reachable set to avoid intersecting with dangerous states. Thereby, we can ensure the safety of the vehicle platooning system. Finally, we present the final algorithm to solve the \text{\textbf{ASAP}} problem.
	
	\subsection{Impacts of Placement and Saturation on Reachable Set}
\begin{figure}
		\centering
		\includegraphics[width=0.8\linewidth]{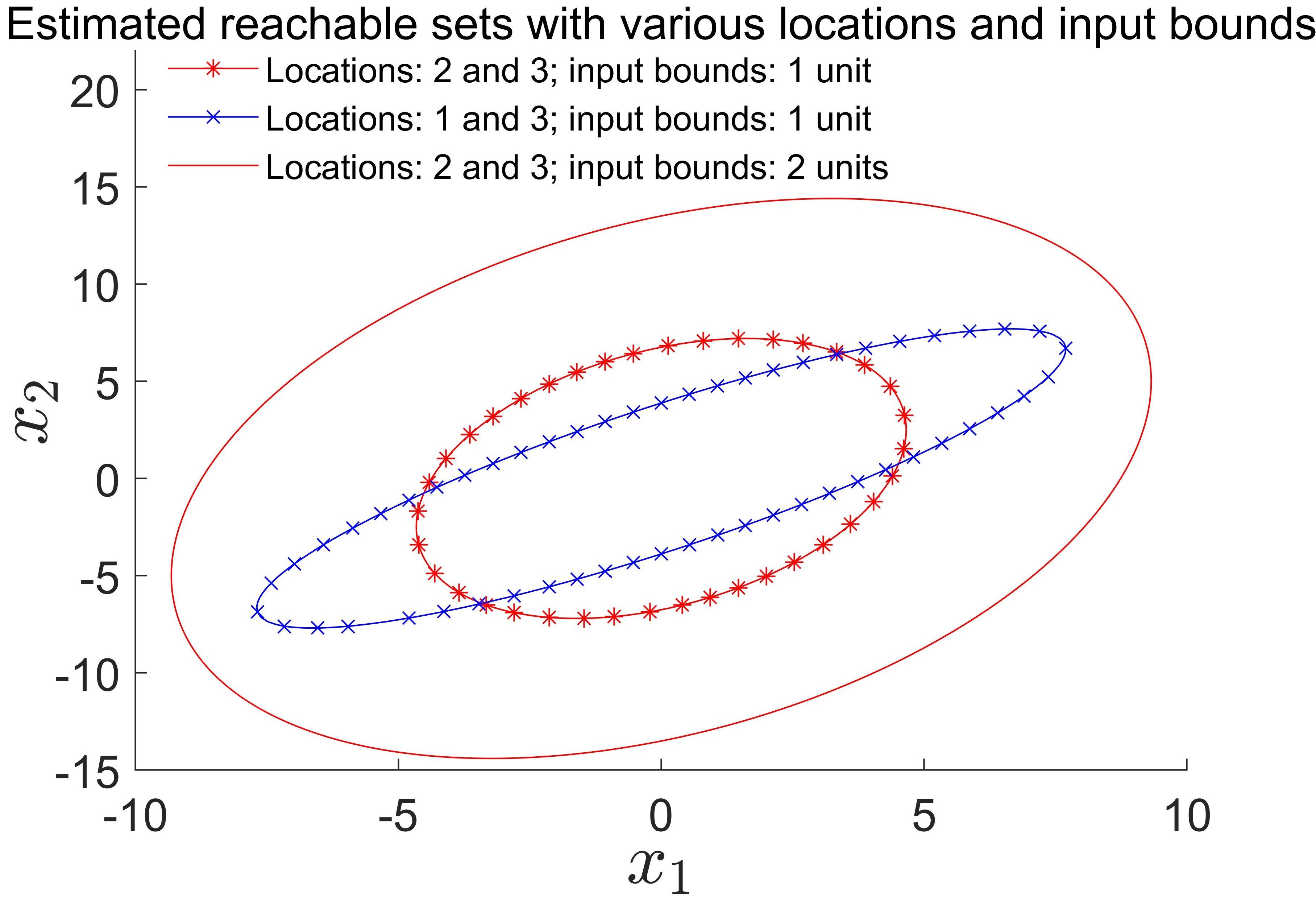}
		\caption{Different ellipsoids encapsulating reachable set (i.e., estimated reachable sets) when selecting different actuator locations and input bounds (only the first two dimensions are shown after projecting a 5-D ellipsoids into a 2-D plane). The detailed simulation results are in Section~\ref{sec:three-vehiclesimulation}.}
		\label{fig:impact}
	\end{figure}

	We note that different combinations of actuator locations and bounds yield different shapes and sizes of estimated reachable sets; see illustrated examples in Fig.~\ref{fig:impact}. 	We can infer from Fig.~\ref{fig:impact} that \textit{(i)} one can effectively control the shape of reachable set $\mathcal{R}_\mathcal{S}$ by strategically choosing the actuator location and \textit{(ii)} adjust the volume of $\mathcal{R}_\mathcal{S}$ by setting actuator bounds, ensuring constraint \eqref{equ:newboundasap} holds. Hence, we first consider the actuator placement problem in Section~\ref{sec:ap}, then 
	followed by solving the actuator saturation problem in Section~\ref{sec:as} with the help of system transformation by the decoupling of locations and bounds in Section~\ref{sec:decoupling}. 
	
	\subsection{Actuator Placement to Determine the Defense Locations} \label{sec:ap}
	The actuator placement problem~\cite{MinimalActuatorPlacement} aims to actuate a small fraction of states of \eqref{eq:LTIS} to optimize system performances. In the vehicle platooning system, it selects the optimal defense locations to maximize the vehicle platooning system's controllability to mitigate attacks. That is, if we do not consider the actuation bounds $\hat{\m \Gamma}$ or equivalently consider the actuation bounds as unit-energy inputs, and the objective function $g(\mathcal{S},\hat{\m \Gamma})$ turns into $g(\mathcal{S};\hat{\m \Gamma})$. Therefore, the actuator placement problem can be written as
	\begin{equation}\label{eq:app}
		\text{\textbf{AP:}} \; \;  \underset{\mathcal{S}}{\text{min}}  \; \; g(\mathcal{S};  \hat{\m \Gamma}) \; \;
		\text{s. t.}  
		\; \;   \mathcal{S} \subseteq \mathcal{V}, |\mathcal{S}| \leq m,
	\end{equation} 
	where the system performance $g(\mathcal{S}; \hat{\m \Gamma})$ is a set function defined as $\log\det(\m W_c^{-1})$ and $\m W_c$ is the controllability Gramian~\cite{boyd1994linear} of \eqref{eq:LTIS}, that is $\m W_c = \sum\limits_{i=0}^{\infty} \m A^i \m B_\mathcal{S}  (\m B_\mathcal{S}  )^\top(\m A^\top)^i$.
	
	Inspired by the fact for a controllable LTI system with unit-energy inputs, the reachable set $\mathcal{R}_\mathcal{S}$ can be encapsulated by an ellipsoid $ \{ \m  x  |\ \m  x^{\top} \m W_c^{-1} \m  x \leq \alpha  \}$~\cite{boyd1994linear}. Note that  a standard ellipsoid $\mathcal{E}(\m P, \alpha)$ is defined as $
	\mathcal{E}(\m P,\alpha):=\left\{ \m  x \in \mathbb{R}^n\ |\ \m  x^{\top} \m P \m  x \leq \alpha \right\}$, where $\m P \in \mathbb{R}^{n \times n}\succ{0} $ and $\alpha \in \mathbb{R}_{>0}$ are the shape matrix and a positive constant of an ellipsoid. When $\alpha=1$, we omit writing it out explicitly, i.e., $\mathcal{E}(\m P,1) = \mathcal{E}(\m P)$. Besides that, the objective function $g(\mathcal{S};\hat{\m \Gamma}) = \log\det(\m W_c^{-1})$ has some special properties that we can take advantage of, and we give a conjecture directly as follows without proof. %
	
	\begin{conjecture}[Supermodularity of $\log\det(\m W_c^{-1})$~\cite{MinimalActuatorPlacement}]
		Consider the discrete-time LTI system~\eqref{eq:LTIS} with system matrices $(\m A, \m B_{\mathcal{S}})$ and individually-bounded control inputs ${\m u}_\mathcal{S}$, the function $\log\det(\m W_c^{-1})$ of the minimum-volume ellipsoid $V(\mathcal{E}_{min})$ which encapsulates the entire reachable set $\mathcal{R}_\mathcal{S}$ is supermodular with respect to the choice of $\mathcal{S}$.
	\end{conjecture}
	
	We note that a set function $g(\cdot)$ is supermodular if and only if $ g(\mathcal{S} \cup\{e\})-g(\mathcal{S}) \leq g(\mathcal{S}' \cup\{e\})-g(\mathcal{S}')$ for any subsets $\mathcal{S} \subseteq \mathcal{S}' \subseteq \mathcal{V}$ and $\{e\} \in \mathcal{V} \backslash \mathcal{S}'$~\cite{TahaRevisit,MinimalActuatorPlacement}. A set function $g(\cdot)$ is submodular if $-g(\cdot)$ is supermodular. Intuitively, submodularity is a diminishing returns property where adding an element $e$ to a smaller set gives a larger gain than adding one to a larger set~\cite{lovasz1983submodular,TahaRevisit}.
	
	The supermodularity framework in set function optimization allows greedy algorithms, which are computationally efficient and perform well despite being suboptimal. Notably, the set function $g(\mathcal{S})$  is supermodular~\cite{MinimalActuatorPlacement}, making it suitable for a classic greedy algorithm that can solve the NP-hard problem \text{\textbf{AP}}  and yield a solution $\mathcal{S}$ with at least 63\% of the optimal value $g(\mathcal{S}^{*})$~\cite{Tzoumas2016}. Empirical studies~\cite{Tzoumas2016,Zhang2017,Cortesi2014,TahaRevisit} suggest that greedy algorithms can often achieve near-optimal solutions, not just 63\% of the optimal.
	
	\subsection{Actuator Saturation to Redesign Bounds}\label{sec:as}
	%With the optimal defense location $\mathcal{S}^{*}$ of \textbf{AP} problem, that is, the zeros and ones are fixed in ${\m B}_\mathcal{S}$, and the shape of $\mathcal{R}_\mathcal{S}$ representing all possible distances and velocities the vehicle platooning system states can reach is determined. This section, we consider the actuator saturation problem~\cite{DARIA, ActuatorSaturation, ConstrainingAttackers} aiming to guarantee that the system would avoid collisions or over-speeds denoted as the dangerous set $\mathcal{D}$ by scaling the $\mathcal{R}_\mathcal{S}$, i.e., redesigning new input bounds $\hat{\m \Gamma}$. 
	
	%We now focus on the actuator saturation (\text{\textbf{AS}}) problem. The goal is to redesign the input bounds to ensure that the ellipsoid representing the reachable set does not intersect with the dangerous set $\mathcal{D}$. To achieve this, we formulate the \text{\textbf{AS}} problem as an optimization problem that seeks to minimize the trace of the redesigned bounds $\hat{\m \Gamma}$, subject to constraints that ensure the redesigned ellipsoid does not intersect with $\mathcal{D}$. 
	
	With the optimal defense location  $\mathcal{S}^{*}$ for the \textbf{AP} problem fixed in ${\m B}_\mathcal{S}$, the reachable set $\mathcal{R}_\mathcal{S}$ defines the distances and velocities the vehicle platooning system can reach. This section addresses the (\text{\textbf{AS}}) problem~\cite{DARIA, ActuatorSaturation, ConstrainingAttackers}, aiming to prevent collisions or excessive speeds---represented by the dangerous set $\mathcal{D}$ ---by redesigning the input bounds $\hat{\m \Gamma}$. Specifically, we frame the \textbf{AS} problem as an optimization task to minimize the trace of the new bounds $\hat{\m \Gamma}$ while ensuring the redesigned ellipsoid for the reachable set does not intersect with  $\mathcal{D}$.

	Note that although the true reachable set $\mathcal{R}_\mathcal{S}$ of \eqref{eq:LTIS} is cumbersome to formulate, we can find a minimum-volume ellipsoid $\mathcal{E}(\m P, \alpha)$ with $\m P$ and $\alpha$ as the shape matrix and center encapsulating reachable set. The ellipsoid $\mathcal{E}(\m P, \alpha)$ is a kind of over-approximations of the reachable set (i.e., $\overline{\mathcal{R}}_\mathcal{S}$). As long as $\mathcal{E}(\m P, \alpha) \cap \mathcal{D} = \emptyset$, the constraint \eqref{equ:safety} can be satisfied. 
	
	Hence, to compute the distance between the ellipsoid $\mathcal{E}(\m P, \alpha)$ and a single hyperplane $\m c^\top \m x = b$ (which represents part of the dangerous set $\mathcal{D}$), the distance $d$ is given by $d = \frac{|b| - \sqrt{\alpha \m c^\top \m P^{-1} \m c}}{\sqrt{\m c^\top \m c}}$~\cite{ActuatorSaturation}. The critical condition occurs when the ellipsoid is tangent to the hyperplane, which implies $d \geq 0$. This leads to the following inequality  $\m c^\top \m Y \m c \leq \frac{b^2}{\alpha}$, where $\m Y = \m P^{-1}$ is positive definite.
	
	We adopt the results in~\cite{ActuatorSaturation, ConstrainingAttackers} to find the ellipsoid. Consider LTI system \eqref{eq:LTIS} with known matrices $(\m A,\ \m B_\mathcal{S})$, the original actuator bounds $\gamma_i>0$, $i=1,\dots,m$ collected in matrix $\m \Gamma$, and a set $\mathcal{D}$ of dangerous states bounded by the hyperplanes as~\eqref{eq:danger}. For given $a\in(0,1)$, if there exists a positive definite matrix $\m Y \in \mathbb{R}^{(2n-1) \times (2n-1)}$ and diagonal matrix $\hat{\m \Gamma}:= \diag(\hat{r}_1,\ldots,\hat{r}_m) \in \mathbb{R}^{m \times m}$, $\hat{r}_i>0$, then the new ellipsoid $\mathcal{E}(\m Y^{-1},m)$ encapsulating safe reachable set can be determined by solving the following \textbf{AS} problem
	\begin{subequations} \label{eq:ASP}
		\begin{align}
			\text{\textbf{AS:}} \; \;  \;\underset{\hat{\m \Gamma},\m Y}{\text{min}}  & \; \;  \;  \Tr(\hat{\m \Gamma}) \\
			\text{s. t.}  &\; \; \; \; \;\hat{\m \Gamma} \geq \m \Gamma, \\
			&\; \; \; \; \;  \m Y \succ 0, \label{eq:pos1} \\
			&\quad\ \m c_i^{\top} \m Y \m c_i \leq \frac{\m b_i^2}{m},\quad \text{for  }i=1\ldots,\kappa, \label{eq:pla}\\
			&\quad \begin{bmatrix}
				a \m Y & 0 & \m Y\m A^{\top}\\
				0 & (1-a)\hat{\m \Gamma} & \m B^{\top}_{\mathcal{S}}\\
				\m A \m Y & \m B_{\mathcal{S}} & \m Y
			\end{bmatrix} \succeq 0 \label{eq:pos2}, 
		\end{align} 
	\end{subequations}
	where the objective function $\Tr(\hat{\m \Gamma})$ is minimized to maximize the volume of the new minimum-volume ellipsoid $\mathcal{E}(\m Y^{-1},m)$. Thus, the {new actuator bounds} $\hat{\gamma_i}:=(1/\hat{r}_i)$, $i=1,\dots,m$, enforce that $\mathcal{E}(\m Y^{-1},m) \cap \mathcal{D} = \emptyset$ that are implemented by constraints~\eqref{eq:pos1},~\eqref{eq:pla} and~\eqref{eq:pos2}. For derivation details, please refer to~\cite{ActuatorSaturation, ConstrainingAttackers}.  
	
	In summary, the \text{\textbf{ASAP}} problem is solved by determining the optimal actuator defense location $\mathcal{S}$ in \text{\textbf{AP}} problem~\eqref{eq:app} and re-designing the actuator bounds $\hat{\m \Gamma}$ in \text{\textbf{AS}} problem~\eqref{eq:ASP}. We give the detailed algorithm next.
	
	\subsection{A Two-stage ASAP Algorithm}
	
	This section presents a two-stage ASAP algorithm (i.e., Algorithm~\ref{alg:1}) to constrain the attacker's capabilities by considering the actuator locations and bounds jointly. AP Stage of Algorithm~\ref{alg:1} selects the most controllable locations as defense positions; AS Stage redesigns the corresponding actuator bounds to avoid dangerous states.
	
	\begin{algorithm}
		\label{alg:1}
		\caption{A Two-stage ASAP Algorithm}
		\small	\DontPrintSemicolon
		\KwData{All locations set $\mathcal{V}$, number of actuator $m$, system matrix $\m A$, original input bounds $\m \Gamma$, dangerous set $\mathcal{D}$, constant $a\in(0,\ 1)$} \;
		\KwResult{Selected actuator location set $\mathcal{S}$ and its corresponding re-designed input bounds $\m{\hat{\Gamma}}$} \;
		\textcolor{blue}{// \textbf{AP Stage: Determine actuator locations} $\mathcal{S}$}		\;
		Initialize $i = 1, \mathcal{S} =  \emptyset $\;
		\While {$i \leq m$}{
			$e_{i} \leftarrow  \mathrm{argmax} _{e  \in \mathcal{V} \backslash \mathcal{S} }\left[g(\mathcal{S})-g(\mathcal{S}  \cup\{e\})\right]$ \;
			$\mathcal{S}  \leftarrow \mathcal{S}  \cup\left\{e_{i}\right\}$\;
			$i \leftarrow i+1$		
		}
		Let $\mathcal{S}^{*} \leftarrow \mathcal{S}$,  actuator location set $\mathcal{S}^{*}$ is obtained\;
		\textcolor{blue}{// \textbf{AS Stage: Re-design actuator bounds } $\m{\hat{\Gamma}}$}\;
		Construct input matrix $\m{B}_{\mathcal{S}}$ according to solved $\mathcal{S}^{*}$\;
		Initialize $a = 0$, $\Delta a = 0.1$, volume $V_{\min} = \infty$, $\m{\hat{\Gamma}}_{\min} = \m 0$\;
		\While{$a <1$}{
			Solve \eqref{eq:ASP} to obtain $\m Y$, $\hat{\m \Gamma}$ \;
			Calculate the current volume $V(\mathcal{E}(\m Y^{-1},m))$\;
			\If{$V(\mathcal{E}(\m Y^{-1},m)) < V_{\min}$}{
				$V_{\min} \leftarrow V(\mathcal{E}(\m Y^{-1},m))$,
				$\m{\hat{\Gamma}}_{\min} \leftarrow {\hat{\m \Gamma}}$\;
			}
			$a \leftarrow a + \Delta a$
		}
		$\m{\hat{\Gamma}} \leftarrow {\hat{\m \Gamma}}_{\min}  $
	\end{algorithm}

	In AP Stage, i.e., Lines 4 to 9 in Algorithm~\ref{alg:1}, the actuator set $\mathcal{S}$ is set to empty. Then, a loop is initiated, and an actuator location element $e_i$ is selected from the set in $\mathcal{V}$ but not in $\mathcal{S}$. The selection is based on maximizing the difference between the current set $\mathcal{S}$ and the set $\mathcal{S}\cup e$, and the $g(\mathcal{S})$ is chosen as a supermodular function $\log\det(\m W_c^{-1})$ as mentioned in Section~\ref{sec:ap}. This ensures that the selected actuator provides the maximum controllability increment to the system. Note that when system is not controllable (i.e., $\m W_c^{-1}$ does not exist), the $g(\mathcal{S})$ can be updated as $\log\det(\m W_c + \epsilon \m I)^{-1}$, and $\epsilon$ is a constant; see~\cite{MinimalActuatorPlacement} for details. Once $e_i$ is selected, it is added to the actuator set $\mathcal{S}$. The counter $i$ is then incremented by 1 to proceed to the next iteration. The loop continues until $i$ exceeds $m$, the total number of actuators available. After the loop completes, the sub-optimal actuator set $\mathcal{S}^{*}$ is determined.

	In AS Stage, i.e., Lines 11 to 21 in Algorithm~\ref{alg:1}, the input matrix $\m B_\mathcal{S}$ is constructed based on $\mathcal{S}^{*}$, and the parameters are initialized first. A loop is initiated, where $a$ varies from 0 to 1. Within this loop, the \text{\textbf{AS}} problem \eqref{eq:ASP} is solved to obtain $\m Y$ and $\hat{\m \Gamma}$. The corresponding volume $V(\mathcal{E}(\m Y^{-1},m))$ is computed and compared to the minimum volume of the tightest ellipsoid encapsulating reachable set. The loop continues until $a$ reaches 1, and when the tightest ellipsoid is obtained, the corresponding redesigned actuator bound $\hat{\m \Gamma}$ is also determined.

	\section{Case Study} \label{sec:casestudy}
	\subsection{Setups and Validation Method} \label{sec:setups}
	
	\textit{1)} {Setups:} Considering an $n$-vehicle scenario in Fig.~\ref{fig:VP} and using the CACC strategy, and sample interval  $\Delta t = 0.5$  seconds, desired relative distance and velocity $d^*_i = 2$ and $v^*_i = 60$ km/h. The proportional and derivative gains are $k_i^p = 0.2$ and $k_i^d = 0.3$. Additionally, the velocity loss due to friction is set to -0.1. The system model is described by~\eqref{eq:LTIS}. Moreover, the dangerous set~\eqref{eq:danger} to avoid collisions in the platooning scenario is
	\begin{equation}~\label{eq:dangerset}
		\mathcal{D}=\{  \cup_{i=1}^{n-1}   d_i\leq 0 \} = \{  \  \cup_{i=1}^{n-1} \m x_i\leq -d_i^* \}.
	\end{equation}
	
	\textit{2)} {Validation Method:} To validate the effectiveness of the proposed method, we compare two defense strategies: \textit{(i)} selecting all actuators with unit bounds and \textit{(ii)} selecting optimal actuators with redesigned bounds $\hat{\m \Gamma}$. Specifically, the ellipsoid $\mathcal{E}^{\mathcal{V}}$ considering placing actuators at all locations $\mathcal{V}$ with unit bounds (i.e., using $\m B _\mathcal{V}$) is computed as the original over-approximations of reachable set. In contrast, the ellipsoid $\mathcal{E}^{\mathcal{S*}}$ considering placing actuators at selected locations $\mathcal{S}^{*}$ with redesigned bounds $\hat{\m \Gamma}$ (i.e., using $\m B_\mathcal{S} \m u_\mathcal{S}$) is computed as the final over-approximations of reachable set that is not overlapping with dangerous set $\mathcal{D}$. The effectiveness of proposed method is verified through $\mathcal{E}^{\mathcal{V}} \cap \mathcal{D} \neq \emptyset$, and $\mathcal{E}^{\mathcal{S*}} \cap \mathcal{D} = \emptyset $. 
	%
	%	\begin{table}
		%		\renewcommand\arraystretch{1.1}
		%		\caption{Parameter settings.}
		%		\centering
		%		\begin{tabular}{c|c|c}
			%			\hline
			%			Sample interval&  $\Delta t$& $0.5$ s\\
			%			\hline
			%			Desired relative  distance&  $d^*_i$, $i = 1,\ldots,n-1$& $2$ m\\
			%			\hline
			%			Desired velocity&  $v^*_i$, $i = 1,\ldots,n$& $60$ km/h\\
			%			\hline
			%			Proportional gains&  $k_i^p$, $i = 1,\ldots,n$& 0.2\\
			%			\hline
			%			Derivative gains&  $k_i^d$, $i = 1,\ldots,n$& 0.3\\
			%			\hline
			%			Velocity loss caused by friction&  $\beta_i$, $i = 1,\ldots,n$& -0.1\\
			%			\hline
			%			\hline
			%		\end{tabular}
		%		\label{tab:parameter set}
		%               \vspace{-20pt}
		%	\end{table}
	
	Next, we present a simple three-vehicle platooning system to help readers intuitively understand our proposed method. Then, we present a relatively large-scale system to present our results further.
	\subsection{Three-vehicle Platooning System} ~\label{sec:three-vehiclesimulation}

\begin{figure}
		\centering
		\includegraphics[width=0.8\linewidth]{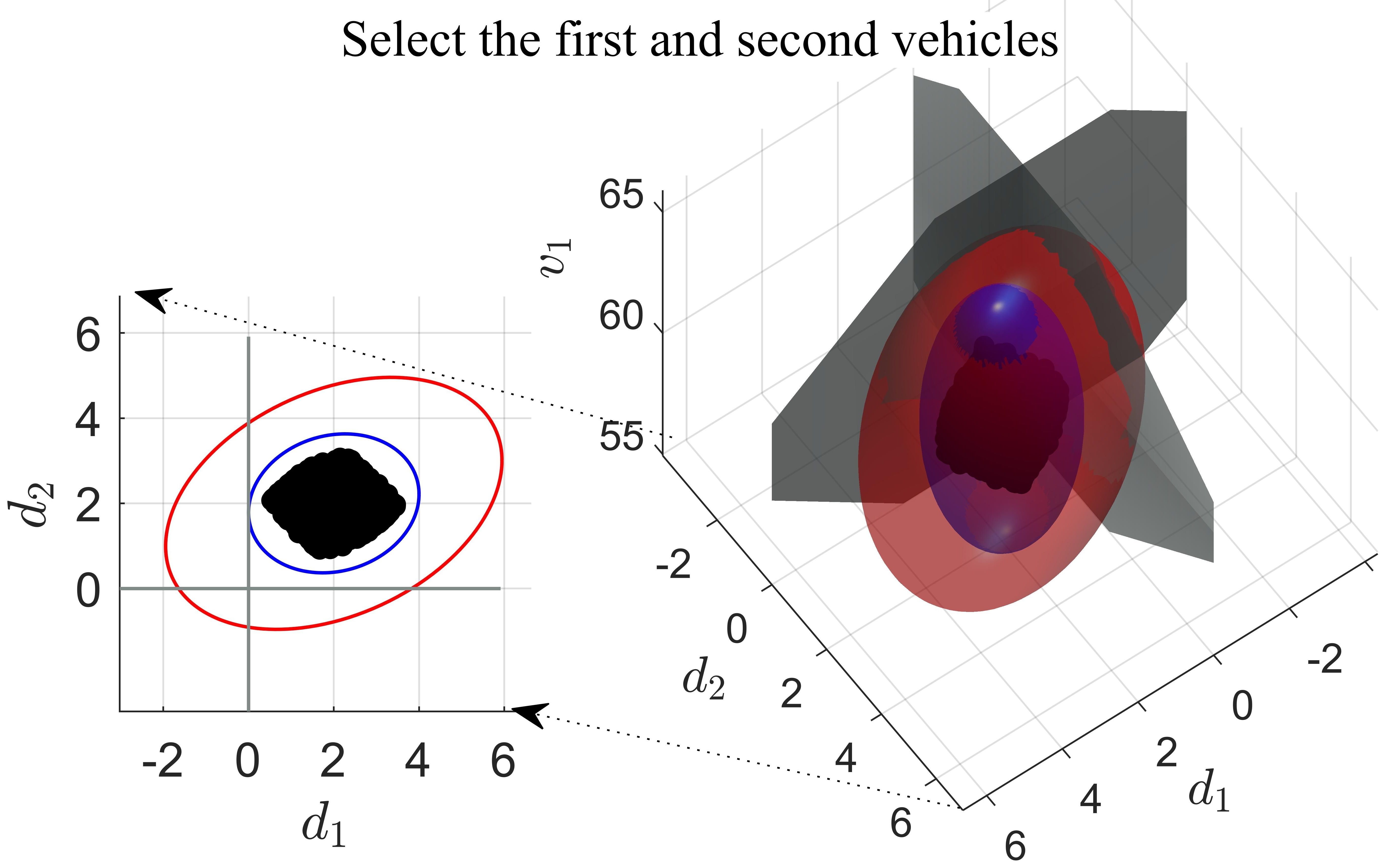}
		\caption{The 2-D and 3-D projections of two 5-D ellipsoids (red original $\mathcal{E}^{\mathcal{V}}$ before applying ASAP algorithm; blue final $\mathcal{E}^{\mathcal{S*}}$ after applying ASAP algorithm), two hyperplanes (gray; dangerous set $\mathcal{D}$), and the empirical reachable set (black; Monte Carlo simulation) for the three-vehicle platooning system when the number of actuators $m = 2$.
		}
		\label{fig:three cars}
	\end{figure}

	After setting up $n=3$ and using the parameters in Section~\ref{sec:setups}, the corresponding matrices in~\eqref{equ:systemmatrix}, i.e., $\m A \in \mathbb{R}^{5\times5}$ and $\m B_{\mathcal{V}} \in \mathbb{R}^{5\times 3}$ are obtained. Besides that, in a three-vehicle platooning system, all locations set $\mathcal{V} = \{1, 2, 3\}$,  original input bounds $\m \Gamma = \m I_3$ (i.e., unit bound is set, and $|  u_i | \leq \sqrt{\gamma_i} = 1, i = 1,2,3$),  and the dangerous set is the region formed by two hyperplanes $\mathcal{D}  =\{d_1 \leq 0 \cup d_2 \leq 0 \}$ according to \eqref{eq:dangerset}; see the two gray hyperplanes in Fig.~\ref{fig:three cars}.
	
	When we set the number of actuators as $m = 2$ and run Algorithm~\ref{alg:1}, the actuator placement result is to select the first and second vehicles that are the most controllable locations to defense, that is, $\mathcal{S}^* = \{1, 2\}$, and the redesigned actuator bounds $\hat{\m \Gamma} = \diag(4.5459,2.1787)$ (i.e., $|  u_1 | \leq \sqrt{\gamma_1} = 0.4690$ and $|  u_2 | \leq \sqrt{\gamma_2} = 0.6775$ in Table~\ref{tab:sensor_impact}). The corresponding results are visualized in Fig.~\ref{fig:three cars}. Note that the  $\mathcal{E}^{\mathcal{V}}$ is a 5-dimensional (5-D) ellipsoid, and the red ellipsoid is the projection of $\mathcal{E}^{\mathcal{V}}$ in a 3-D space. This also applies to $\mathcal{E}^{\mathcal{S*}}$ (blue ellipsoid) and two hyperplanes (gray planes). Interestingly, the blue ellipsoid is not contained in the red ellipsoid because there is no constraint in the $v_1$ direction, making the vehicle platooning system flexible.
	
	We continue to make the results in 2-D space, and it is clear that $\mathcal{E}^{\mathcal{V}} \cap \mathcal{D} \neq \emptyset $ and $\mathcal{E}^{\mathcal{S*}} \cap \mathcal{D} = \emptyset$, i.e., the blue 2-D ellipsoid is tangent to the gray line $d_1 = 0$. In addition, the correctness of the estimated reachable set $\mathcal{E}^{\mathcal{S*}}$ is verified by Monte Carlo simulation (i.e., the black area consisting of black scatted dots is the empirical true reachable set $\mathcal{R}_\mathcal{S}$).

	We also find out that various locations bring various ellipsoid shapes, and once the location is fixed, the size of ellipsoids scales along with the control input bounds; see Fig.~\ref{fig:impact} for details. In addition,  when $m = 3$, that is all three locations are selected to defense attackers, the redesigned actuator bounds are $|  u_1 | \leq \sqrt{\gamma_1} = 0.3404, |  u_2 | \leq \sqrt{\gamma_2} = 0.4464, |  u_3 | \leq \sqrt{\gamma_3} = 0.3404$. The corresponding 2-D and 3-D projections are omitted due to space limitations. It is intuitive that the more actuator locations are selected, the lower the defense efficiency. This phenomenon is evident in a relatively large-scale system such as the 20-vehicle platooning system introduced next.

	\subsection{20-vehicle Platooning System}

\begin{figure}[h]
	\centering
	\includegraphics[width=0.9\linewidth]{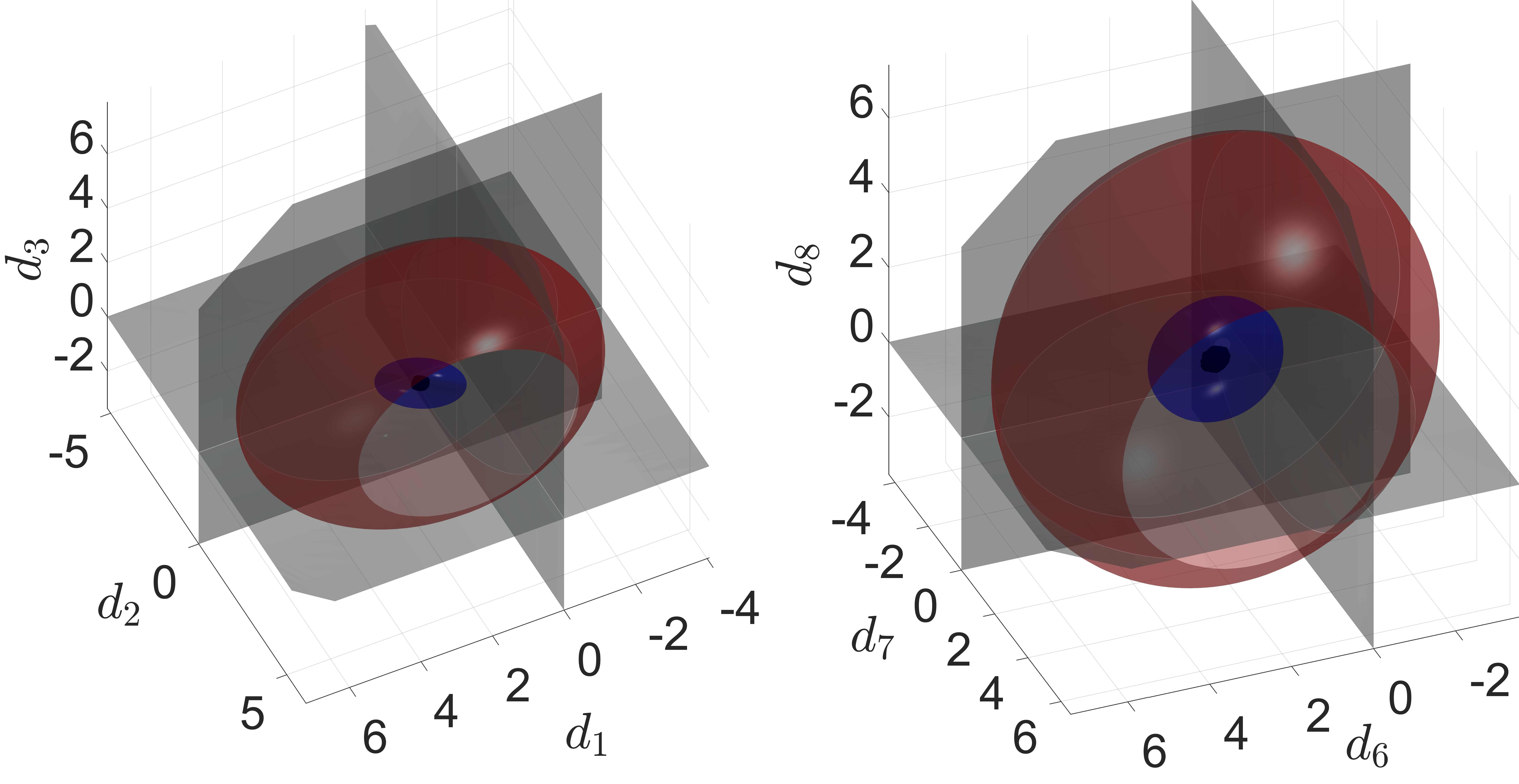}
	\caption{The 3-D projections on $d_1-d_2-d_3$ and $d_6-d_7-d_8$ subspaces of 39-D ellipsoids (red original $\mathcal{E}^{\mathcal{V}}$ is cut to show blue final $\mathcal{E}^{\mathcal{S*}}$ clearly), two hyperplanes (gray; dangerous set $\mathcal{D}$), and the empirical reachable set (black; Monte Carlo simulation) for the 20-vehicle platooning system when the number of actuators $m = 12$.}
	\label{fig:v20}
\end{figure}

	Similarly, after setting up $n=20$, we obtain a 20-vehicle platooning system with $\m x \in \mathbb{R}^{39}$ indicating the obtained ellipsoids encapsulating reachable sets are also in $\mathbb{R}^{39}$. The final results are shown in Table~\ref{tab:sensor_impact} with $m = 8,9,12$. It is clear that the $\mathcal{S}^{*}_{m=12} = \mathcal{S}^{*}_{m=9} \cup \{5,11,16\}$ and $\mathcal{S}^{*}_{m=9} = \mathcal{S}^{*}_{m=8} \cup \{8\}$ thereby showcasing the greedy-optimal solution for Algorithm~\ref{alg:1} that exploits the supermodularity of the set function optimization in Section~\ref{sec:ap}. The ranges of redesigned bounds ensuring the ellipsoids $\mathcal{E}^{\mathcal{S*}}$ not intersecting the dangerous set are decreasing along with the increasing number of selected actuators. When $m=12$, the 3-D projections on $d_1-d_2-d_3$ and $d_6-d_7-d_8$ subspaces of 39-D ellipsoids are shown in Fig.~\ref{fig:v20} to shown the effectiveness of the proposed method. Defense efficiency is ensured by maintaining safety while using only 8, 9, or 12 defense locations, a minimal resource compared to all 20.

	\begin{table}[h]
		\fontsize{7}{7}\selectfont
		\centering
		\setlength\tabcolsep{1pt}
		\renewcommand{\arraystretch}{1.5}
		\makegapedcells
		\setcellgapes{1.0pt}
		\caption{Actuator placement and saturation results.} ~\label{tab:sensor_impact}
		\begin{tabular}{c|c|c}
			\hline
			\textit{Scenario} & $m$ & \textit{Result (not selected actuator bounds are zeros by default)} \\ \hline
			\multirow{2}{*}{\makecell{  \textit{Three-vehicle}\\ \textit{platooning}\\
					\textit{system}}} & $2$& \makecell{$\sqrt{\gamma_{\textcolor{black}{1}}} = 0.4690,$ $\sqrt{\gamma_{\textcolor{black}{2}}} = 0.6775$ }\\ \cline{2-3} 
			& $3$ & \makecell{$\sqrt{\gamma_{\textcolor{black}{1}}} = 0.3404,$ $\sqrt{\gamma_{\textcolor{black}{2}}} = 0.4464,$ $\sqrt{\gamma_{\textcolor{black}{3}}} = 0.3404$ }\\ \hline
			\multirow{3}{*}{
				\makecell{\\
					\\
					\\  \textit{20-vehicle}\\ 
					\textit{platooning}\\
					\textit{system}}
			}
			& $8$& \makecell{$\sqrt{\gamma_{\textcolor{black}{2}}} = 0.1525,$
				$\sqrt{\gamma_{\textcolor{black}{4}}} = 0.1596,$
				$\sqrt{\gamma_{\textcolor{black}{7}}} = 0.1511,$\\
				$\sqrt{\gamma_{\textcolor{black}{10}}} = 0.1436,$
				$\sqrt{\gamma_{\textcolor{black}{13}}} = 0.1191,$
				$\sqrt{\gamma_{\textcolor{black}{14}}} = 0.1186,$\\
				$\sqrt{\gamma_{\textcolor{black}{17}}} = 0.1433,$
				$\sqrt{\gamma_{\textcolor{black}{19}}} = 0.1380$} \\ \cline{2-3} 
			& $9$& \makecell{$\sqrt{\gamma_{\textcolor{black}{2}}} = 0.1200,$
				$\sqrt{\gamma_{\textcolor{black}{4}}} = 0.1265,$
				$\sqrt{\gamma_{\textcolor{black}{7}}} = 0.1199,$\\
				$\sqrt{\gamma_{\textcolor{black}{8}}} = 0.1197,$
				$\sqrt{\gamma_{\textcolor{black}{10}}} =  0.1272,$
				$\sqrt{\gamma_{\textcolor{black}{13}}} = 0.1237,$\\
				$\sqrt{\gamma_{\textcolor{black}{14}}} = 0.1224,$
				$\sqrt{\gamma_{\textcolor{black}{17}}} = 0.1281,$
				$\sqrt{\gamma_{\textcolor{black}{19}}} = 0.1214$} \\ \cline{2-3} 
			& $12$& \makecell{
				$\sqrt{\gamma_{\textcolor{black}{2}}} = 0.0972,$
				$\sqrt{\gamma_{\textcolor{black}{4}}} = 0.1015,$
				$\sqrt{\gamma_{\textcolor{black}{5}}} = 0.1019,$\\
				$\sqrt{\gamma_{\textcolor{black}{7}}} = 0.0926,$
				$\sqrt{\gamma_{\textcolor{black}{8}}} = 0.0911,$
				$\sqrt{\gamma_{\textcolor{black}{10}}} =  0.0841,$\\
				$\sqrt{\gamma_{\textcolor{black}{11}}} =  0.0841,$
				$\sqrt{\gamma_{\textcolor{black}{13}}} = 0.0911,$
				$\sqrt{\gamma_{\textcolor{black}{14}}} =  0.0926,$\\
				$\sqrt{\gamma_{\textcolor{black}{16}}} = 0.1019,$
				$\sqrt{\gamma_{\textcolor{black}{17}}} = 0.1015,$
				$\sqrt{\gamma_{\textcolor{black}{19}}} = 0.0972$
			}
			\\ \hline \hline
		\end{tabular}%
		%  	}
	\vspace{-10pt}
\end{table}

\section{Conclusion}\label{sec:conclusion}
%The main objective of this paper is to redesign the actuator locations and bounds to enhance the safety and resilience of vehicle platooning systems. We present a new formulation/model integrating the actuator placement and saturation. We propose a systematic and efficient two-stage algorithm method to co-optimize the actuator locations and bounds and illustrate its effectiveness through numerical simulations. Our future work will focus on enhancing control efficiency.

This paper addresses maintaining safety and defense efficiency in vehicle platooning systems under potential cyber-attacks. By focusing on control node placement and optimizing input bounds for key actuators, we develop a defense strategy that limits the impact of attacks while preserving the system's ability to operate effectively. Rather than restricting all actuators with low defense efficiency, we demonstrate that selectively constraining only critical actuators enhances safety. Our analysis of the system's reachable set under state and input bounds shows that this approach provides a practical balance between safety and defense efficiency in vehicle platooning. Future work could explore the real-world implementation of these strategies and extend the approach to more complex multi-lane systems.

\bibliographystyle{IEEEtran}
\bibliography{IEEEabrv,reference.bib}

\begin{thebibliography}{10}
\providecommand{\url}[1]{#1}
\csname url@rmstyle\endcsname
\providecommand{\newblock}{\relax}
\providecommand{\bibinfo}[2]{#2}
\providecommand\BIBentrySTDinterwordspacing{\spaceskip=0pt\relax}
\providecommand\BIBentryALTinterwordstretchfactor{4}
\providecommand\BIBentryALTinterwordspacing{\spaceskip=\fontdimen2\font plus
\BIBentryALTinterwordstretchfactor\fontdimen3\font minus
  \fontdimen4\font\relax}
\providecommand\BIBforeignlanguage[2]{{%
\expandafter\ifx\csname l@#1\endcsname\relax
\typeout{** WARNING: IEEEtran.bst: No hyphenation pattern has been}%
\typeout{** loaded for the language `#1'. Using the pattern for}%
\typeout{** the default language instead.}%
\else
\language=\csname l@#1\endcsname
\fi
#2}}

\bibitem{goodplatoon}
S.~C. Calvert, G.~Mecacci, D.~D. Heikoop, and F.~S. de~Sio, ``Full platoon
  control in truck platooning: A meaningful human control perspective,'' in
  \emph{2018 21st International Conference on Intelligent Transportation
  Systems (ITSC)}, 2018, pp. 3320--3326.

\bibitem{CACCreview}
M.~A. Tahiri, A.~Rachid, B.~Boudmane, I.~Mortabit, and S.~Laaroussi, ``Toward
  cooperative adaptive cruise control: A mini-review,'' in \emph{2024
  International Conference on Circuit, Systems and Communication (ICCSC)},
  2024, pp. 1--6.

\bibitem{defendplatoon1}
G.~Sun, T.~Alpcan, B.~I.~P. Rubinstein, and S.~Camtepe, ``To act or not to act:
  An adversarial game for securing vehicle platoons,'' \emph{IEEE Transactions
  on Information Forensics and Security}, vol.~19, pp. 163--177, 2024.

\bibitem{maninthemiddle}
M.~Thankappan, H.~Rifà-Pous, and C.~Garrigues, ``A signature-based wireless
  intrusion detection system framework for multi-channel man-in-the-middle
  attacks against protected wi-fi networks,'' \emph{IEEE Access}, vol.~12, pp.
  23\,096--23\,121, 2024.

\bibitem{jammingattack}
Y.~Hu, H.~Shan, R.~G. Dutta, and Y.~Jin, ``Protecting platoons from stealthy
  jamming attack,'' in \emph{2020 Asian Hardware Oriented Security and Trust
  Symposium (AsianHOST)}, 2020, pp. 1--6.

\bibitem{Murguia2020}
C.~Murguia, I.~Shames, J.~Ruths, and D.~Nešić, ``Security metrics and
  synthesis of secure control systems,'' \emph{Automatica}, vol. 115, p.
  108757, 5 2020.

\bibitem{ActuatorSaturation}
S.~H. Kafash, J.~Giraldo, C.~Murguia, A.~A. Cardenas, and J.~Ruths,
  ``Constraining attacker capabilities through actuator saturation,'' in
  \emph{2018 Annual American Control Conference (ACC)}.\hskip 1em plus 0.5em
  minus 0.4em\relax IEEE, 6 2018, pp. 986--991.

\bibitem{DARIA}
J.~Giraldo, S.~H. Kafash, J.~Ruths, and A.~A. Cardenas, ``Daria: Designing
  actuators to resist arbitrary attacks against cyber-physical systems,'' in
  \emph{2020 IEEE European Symposium on Security and Privacy}.\hskip 1em plus
  0.5em minus 0.4em\relax IEEE, 9 2020, pp. 339--353.

\bibitem{ConstrainingAttackers}
S.~H. Kafash, N.~Hashemi, C.~Murguia, and J.~Ruths, ``Constraining attackers
  and enabling operators via actuation limits,'' \emph{Proceedings of the IEEE
  Conference on Decision and Control}, vol. 2018-Decem, pp. 4535--4540, 12
  2018.

\bibitem{butchkocps2021}
D.~Butchko, B.~Croteau, and K.~Kiriakidis, ``Cyber-physical system security of
  surface ships using intelligent constraints,'' in \emph{2021 IEEE
  International Conference on Communications Workshops}.\hskip 1em plus 0.5em
  minus 0.4em\relax IEEE, 6 2021, pp. 1--6.

\bibitem{Escudero2021}
C.~Escudero, C.~Murguia, P.~Massioni, and E.~Zamaï, ``Enforcing safety under
  actuator injection attacks through input filtering,'' in \emph{2022 European
  Control Conference (ECC)}, 2022, pp. 1521--1528.

\bibitem{Escudero2023}
C.~Escudero, C.~Murguia, P.~Massioni, and E.~Zamäi, ``Safety-preserving
  filters against stealthy sensor and actuator attacks,'' in \emph{2023 62nd
  IEEE Conference on Decision and Control (CDC)}.\hskip 1em plus 0.5em minus
  0.4em\relax IEEE, 12 2023, pp. 5097--5104.

\bibitem{transportation1}
G.~Ma, P.~R. Pagilla, and S.~Darbha, ``Assessing the safety benefits of cacc+
  based coordination of connected and autonomous vehicle platoons in emergency
  braking scenarios,'' in \emph{2024 IEEE Intelligent Vehicles Symposium (IV)},
  2024, pp. 2248--2254.

\bibitem{transportation2}
J.~Chen, H.~Liang, J.~Li, and Z.~Lv, ``Connected automated vehicle platoon
  control with input saturation and variable time headway strategy,''
  \emph{IEEE Transactions on Intelligent Transportation Systems}, vol.~22,
  no.~8, pp. 4929--4940, 2021.

\bibitem{MinimalActuatorPlacement}
V.~Tzoumas, M.~A. Rahimian, G.~J. Pappas, and A.~Jadbabaie, ``Minimal actuator
  placement with bounds on control effort,'' \emph{IEEE Transactions on Control
  of Network Systems}, vol.~3, no.~1, pp. 67--78, 3 2016.

\bibitem{taha_time-varying_2019}
A.~F. Taha, N.~Gatsis, T.~Summers, and S.~A. Nugroho, ``Time-varying sensor and
  actuator selection for uncertain cyber-physical systems,'' \emph{IEEE
  Transactions on Control of Network Systems}, vol.~6, no.~2, pp. 750--762, 6
  2019.

\bibitem{VehicularPlatooning}
S.~Dadras, R.~M. Gerdes, and R.~Sharma, ``Vehicular platooning in an
  adversarial environment,'' in \emph{Proceedings of the 10th ACM Symposium on
  Information, Computer and Communications Security}, ser. ASIA CCS '15.\hskip
  1em plus 0.5em minus 0.4em\relax New York, NY, USA: Association for Computing
  Machinery, 4 2015, p. 167–178.

\bibitem{inputsaturationinplatoon}
J.~Wu, Y.~Wang, and C.~Yin, ``Curvilinear multilane merging and platooning with
  bounded control in curved road coordinates,'' \emph{IEEE Transactions on
  Vehicular Technology}, vol.~71, no.~2, pp. 1237--1252, 2022.

\bibitem{boyd1994linear}
S.~Boyd, L.~El~Ghaoui, E.~Feron, and V.~Balakrishnan, \emph{Linear matrix
  inequalities in system and control theory}.\hskip 1em plus 0.5em minus
  0.4em\relax SIAM, 1994.

\bibitem{TahaRevisit}
A.~F. Taha, S.~Wang, Y.~Guo, T.~H. Summers, N.~Gatsis, M.~H. Giacomoni, and
  A.~A. Abokifa, ``Revisiting the water quality sensor placement problem:
  Optimizing network observability and state estimation metrics,''
  \emph{Journal of Water Resources Planning and Management}, vol. 147, no.~7,
  2021.

\bibitem{lovasz1983submodular}
L.~Lov{\'a}sz, ``Submodular functions and convexity,'' in \emph{Mathematical
  programming the state of the art}.\hskip 1em plus 0.5em minus 0.4em\relax
  Springer, 1983, pp. 235--257.

\bibitem{Tzoumas2016}
V.~Tzoumas, A.~Jadbabaie, and G.~J. Pappas, ``{Sensor placement for optimal
  Kalman filtering: Fundamental limits, submodularity, and algorithms},''
  \emph{Proceedings of the American Control Conference}, vol. 2016-July, pp.
  191--196, 2016.

\bibitem{Zhang2017}
H.~Zhang, R.~Ayoub, and S.~Sundaram, ``{Sensor selection for Kalman filtering
  of linear dynamical systems: Complexity, limitations and greedy
  algorithms},'' \emph{Automatica}, vol.~78, pp. 202--210, 2017.

\bibitem{Cortesi2014}
F.~L. Cortesi, T.~H. Summers, and J.~Lygeros, ``{Submodularity of energy
  related controllability metrics},'' in \emph{Proceedings of the IEEE
  Conference on Decision and Control}, vol. 2015-Febru, no. February, mar 2014,
  pp. 2883--2888.

\end{thebibliography}

\end{document}